\UseRawInputEncoding
\documentclass[reprint, twocolumn, superscriptaddress]{revtex4-1}
\usepackage{bm, amsmath, amsfonts, amssymb, braket}
\usepackage{subfigure}
\usepackage{color}
\usepackage{graphicx}
\usepackage{dcolumn} % Align table columns on decimal point

\usepackage{rotating} % for 90 degree rotated table and fig

\begin{document}
\title{Nontrivial interband effect: applications to magnetic susceptibility, nonlinear optics, and topological degeneracy pressure}
\author{Nobuyuki Okuma}
\email{okuma@hosi.phys.s.u-tokyo.ac.jp}
\affiliation{%
 Yukawa Institute for Theoretical Physics, Kyoto University, Kyoto 606-8502, Japan
}%

\date{\today}
\begin{abstract}
The interband effect is an important concept both in traditional and modern solid-state physics. In this paper, we present a theory of the nontrivial interband effect, which cannot be removed without breaking given rules.
We define the general nontrivial interband effect by regarding a property of the set of the total bands of a tight-binding Hamiltonian as the triviality.
As examples of the source of the nontrivial interband effect, we consider several topological concepts: stable topological insulator, symmetry-based indicator, fragile topological insulator, and multipole/higher-order topological insulator.
As an application, we calculate the orbital magnetic susceptibility for tight-binding Hamiltonians with topological properties.
In addition, we consider the mechanical properties induced by the nontrivial interband effect.
We define interband-induced degeneracy pressure, which tends to take a negative value, and calculate it for the Chern insulator.
This calculation demonstrates the importance of topological band structures in mechanical properties of solids.
We also discuss the application to nonlinear optics characterized by the polarization difference and
the generalization to interacting systems with entanglement.
The framework of the nontrivial interband effect, which includes topological concepts as subsets, might be useful for finding unexplored concepts.
\end{abstract}
\maketitle
\section{Introduction}
The interband effect, in which multi-band nature affects the motion of electrons in solids, plays important roles in electromagnetic responses.
In the context of traditional solid-state physics, lots of basic quantities related to the interband effect have been investigated.
The orbital magnetism is a typical example of such quantities \cite{fkubo-69,fkubo-70,fukuyama-71,ogata-fukuyama-1,piechon-1,piechon-2,piechon-3,piechon-4}.
In particular, the diamagnetic response is enhanced by the interband effect in semimetallic systems \cite{McClure,fkubo-70,piechon-1,piechon-2,piechon-3,piechon-4,fuseya-ogata-15,Koshino-Ando-07,ogata-fukuyama-3,tateishi,suetsugu} such as bismuth \cite{fkubo-70,fuseya-ogata-15} and graphite/graphene-type compounds \cite{McClure,Koshino-Ando-07,ogata-fukuyama-3}. 
Notably, such electromagnetic responses are present even in the absence of the Fermi surfaces.
In the context of modern solid-state physics, the topological insulator is an interband-induced concept since the topological properties require at least two bands.
The relationship between topological propeties including Berry phase and orbital magnetism has been investigated in several contexts \cite{niu-review,Trifunovic-19}.

In terms of material search, it might be useful to find criterions for the presence of the interband effect. In this paper, we propose the concept of the nontrivial interband effect, which cannot be removed without breaking given rules.
We first define the triviality as a property of the set of the total bands of the tight-binding Hamiltonian.
By considering the cases in which this property is not satisfied, we define the general nontrivial interband effect between the set of occupied bands and that of the unoccupied bands. As examples of the nontriviality, we consider some topological concepts: stable topological insulator \cite{Kane-review,Zhang-review}, symmetry-based-indicator-type topology \cite{kruthoff,po-vishwanath-watanabe,chemistry}, fragile topological insulator \cite{fragile-18}, and multipole/higher-order topological insulator \cite{Benalcazar-17}.
As an application, we calculate the orbital magnetic susceptibility of tight-binding Hamiltonians with topological properties.
In addition, we investigate the mechanical properties induced by the interband effect.
We define interband-induced degeneracy pressure and find that it tends to take a negative value.
We also discuss the application to nonlinear optical responses and
the generalization of the nontrivial interband effect to interacting systems with entanglement.

This paper is organized as follows.
In Sec. \ref{sec2}, we introduce definitions and conventions of several concepts used in this paper, as well as the characterization of the interband effect in terms of the Hilbert spaces of the occupied bands. In Sec. \ref{sec3}, we formulate the general nontrivial interband effect in terms of the decomposability of the tight-binding Hamiltonian. In Sec. \ref{sec4}, we give examples of nontrivial interband effects that come from several topological concepts.
In Sec. \ref{sec5}, we numerically calculate the orbital magnetic susceptibility for each topological concept treated in Sec. \ref{sec4}, as an application of the nontrivial interband effect. In Sec.  \ref{sec6}, we consider the mechanical properties induced by the nontrivial interband effect. We define the interband-induced degeneracy pressure and calculate it for the Chern insulator. In Sec. \ref{sec7} and \ref{sec8}, we 
give discussions about the application to nonlinear optics and
the generalization to interacting systems with entanglement.

\section{Interband effect and Hilbert spaces of occupied bands \label{sec2}}
In this section, we characterize the interband effect in terms of the Hilbert spaces of occupied bands.
\subsection{Definitions and conventions}
We here introduce several notions in the band theory and remark the conventions used in this paper.
We assume that the total system is well approximated by the tight-binding model with finite-range hopping (or exponentially decaying hopping) and without the interaction terms:
\begin{align}
    \hat{H}^{\rm TB}=\sum_{\bm{R},\bm{R}'}\sum_{i,i'}c^\dagger_{\bm{R},i}[H^{\rm TB}]_{(\bm{R},i),(\bm{R}',i')} c_{\bm{R}',i'}, 
\end{align}
where $\bm{R}$ and $i$ denote the unit cell and intracell atomic orbital,
$c$ is the electron creation operator at $(\bm{R},i)$, and $H^{\rm TB}$ is the martix representation of the tight-binding model.
In this paper, We adopt the local atomic orbitals as the basis vectors of the tight-binding Hamiltonian matrix unless otherwise noted.
We also assume the translation invariance and the periodic boundaries.
Owing to these assumptions, one can introduce the Fourier transform.

There are two types of conventions in the Fourier transform.
The first convention is defined as 
\begin{align}
    c^\dagger_{\bm{k},i}=\sum_{\bm{R}}c^\dagger_{\bm{R},i}~e^{i\bm{k}\cdot(\bm{R}+\bm{r}_i)},
\end{align}
where $\bm{r}_i$ is the intracell relative position of the atomic orbital $i$, and $\bm{k}$ is the crystal momentum.
The tight-binding Hamiltonian can be decomposed into the Bloch Hamiltonian matrices:
\begin{align}
    \hat{H}^{\rm TB}=\sum_{\bm{k}}\sum_{i,j}c^\dagger_{\bm{k},i}[H_{\bm{k}}]_{i,j}
    c_{\bm{k},j}.
\end{align}
The energy dispersion and eigenstates of $H_{\bm{k}}$ are given by 
\begin{align}
    H_{\bm{k}}|u_{\bm{k},\alpha}\rangle=E_{\bm{k},a}|u_{\bm{k},\alpha}\rangle,
\end{align}
where $\alpha$ denotes the band index.

The second convention is defined as 
\begin{align}
    \tilde{c}^\dagger_{\bm{k},i}=\sum_{\bm{R}}c^\dagger_{\bm{R},i}~e^{i\bm{k}\cdot\bm{R}}.\label{second}
\end{align}
In this convention, the Fourier transform does not contain the intracell position $\bm{r}_i$ of the orbital $i$, which enables one to define the Bloch Hamiltonian matrix $\tilde{H}_{\bm{k}}$ and its eigenstates $|\tilde{u}_{\bm{k},\alpha}\rangle$ that are periodic in momentum space:
\begin{align}
    &\tilde{H}_{\bm{k}}|\tilde{u}_{\bm{k},\alpha}\rangle=E_{\bm{k},a}|\tilde{u}_{\bm{k},\alpha}\rangle,\notag\\
    &\tilde{H}_{\bm{k}+\bm{G}}=\tilde{H}_{\bm{k}},\notag\\
    &|\tilde{u}_{\bm{k}+\bm{G},\alpha}\rangle=|\tilde{u}_{\bm{k},\alpha}\rangle,
\end{align}
where $\bm{G}$ is an arbitrary reciprocal lattice vector. 
Note that the absence of the information of the intracell relative position $\bm{r}_i$ in the Bloch Hamiltonian should be reminded for the evaluation of physical quantities such as the electric polarization and the current operator.

The above two conventions are related to each other via the following unitary transformation:
\begin{align}
    H_{\bm{k}}=D_{\bm{k}}\tilde{H}_{\bm{k}}D^{\dagger}_{\bm{k}},\label{firstsecond}
\end{align}
where $D_{\bm{k}}$ is a diagonal unitary matrix diag$(\cdots,e^{-i\bm{k}\cdot\bm{r}_i},\cdots)$.
Using the periodicity of $\tilde{H}_{\bm{k}}$, we obtain the periodicity for the first convention:
\begin{align}
    H_{\bm{k}+\bm{G}}=D_{\bm{G}}H_{\bm{k}}D^\dagger_{\bm{G}}.
\end{align}
Since the unitary transformation preserves the eigenvalues, the dispersion relation does not depend on the choice of the Fourier transform, while the periodicity of the states in momentum space does not hold for $D_{\bm{k}}\neq1$ in the first convention.
The first convention is useful for the evaluation of physical quantities that requires the information of intercell position, while the second convention is useful for considering the topological properties that require the periodicity of the target space.
In this paper, we mainly adopt the first convention and distinguish the quantities in the second convention by adding tilde.

\subsection{Semiclassical picture of interband effect}
In the semiclassical picture, the elctronic transport is described by the motion of the wavepacket that moves on momentum space.
Let us consider the wavepacket characterized by the momentum $\bm{k}$ and the band index $\alpha$.
Under the electromagnetic driving, the electron momentum varies in time, and takes another value $\bm{k}'$ after the driving.
If the dynamics is essentially described in one-band picture, the state at $\bm{k}'$ stays in the band $\alpha$.

If the multi-band nature is important, on the other hand, the wavepacket is affected by other bands, and the $precession$ between different states occurs during the momentum change.
If such a precession of electronic states exists between the different bands, we say there is an interband effect.
For the readers familiar with the semiclassical theory,
we here note that this precession is best represented by the commutation-relation term of the quantum kinetic (Boltzmann) equation \cite{Mishchenko,Rammer,culcer-sekine}: 
\begin{align}
    \frac{\partial g_{\bm{k}}}{\partial t}&=i[g_{\bm{k}},H_{\bm{k}}]+({\rm other~terms}),
\end{align}
where $H_{\bm{k}}$ is the Bloch Hamiltonian matrix, and $g_{\bm{k}}$ is the distribution function matrix, which contains all the information of the semi-classical dynamics of the electrons.

Note that the above interband effect does not include the $core$ contribution from the coupling between the atomic orbitals localized on the same atom. In this paper, we treat only the contributions from the inter-atomic electronic motion.

\subsection{Interband effect as distance between momentum-dependent projections}
As we mentioned, the interband effect can be interpreted as the precession of the electronic states during the momentum change.
Thus, the interband effect of some specific band $\alpha$ is characterized as the difference between eigenstates at different momenta $1-|\langle u_{\bm{k},\alpha}|u_{\bm{k}',\alpha}\rangle|^2$.
This idea can be generalized to the interband effect between the set of the occupied bands and that of the unoccupied bands, which is the main target of this paper.
At each momentum, the set of the occupied bands forms a subspace of the total Hilbert space spanned by $|u_{\bm{k},\alpha}\rangle$.
In this case, the interband effect is characterized as the difference between the subspaces at different momenta.
For convenience, we introduce the momentum-dependent projection operator from the total Hilbert space to the occupied bands:
\begin{align}
    P^{\rm(occ)}_{\bm{k}}:=\sum_{\alpha\in\{{\rm occ}\}}P_{\bm{k},\alpha}:=\sum_{\alpha\in\{{\rm occ}\}}
    | u_{\bm{k},\alpha} \rangle\langle u_{\bm{k},\alpha} |,
\end{align}
where $\{{\rm occ}\}$ denotes the set of occupied bands.
Mathematically, the difference between the subspaces is reduced to that between the projection operators onto the subspaces.
By using the projection operators, one can define the Hilbert-Schmidt $inner$ $product$ and the chordal $distance$ between subspaces \cite{Bemrose2017}:
\begin{align}
    \langle P,Q\rangle_{\rm H.S.}&:={\rm Tr} [P^\dagger Q]={\rm Tr} [P Q],\\
    d_{\rm C}(P,Q)&:=\sqrt{M-\langle P,Q\rangle_{\rm H.S.}},
\end{align}
where $P,Q$ are projection operators onto subspaces with the dimension $M$. Note that the set of subspaces with the dimension $M$ of the total Hilbert space with the dimension $N$ can be regarded as the Grassmannian with the above inner product and distance (metric).
In terms of these notions, the interband effect between the set of the occupied bands and that of the unoccupied bands is defined as a phenomenon in which
\begin{align}
    d_{\rm C}(P^{\rm(occ)}_{\bm{k}},P^{\rm(occ)}_{\bm{k}'})\neq0
\end{align}
holds at least for some $\bm{k},\bm{k}'$.
In the language of the semiclassical picture, this definition implies that the precession during the momentum change is not closed in a momentum-independent subspace.

Note that the distance between subspaces at different momenta depends on the convention of the Fourier transform.
If we adopt the second convention (\ref{second}), the subspace at $\bm{k}+\bm{G}$ can always be identified with that at $\bm{k}$, which is useful for the topological consideration. 
However, the distance defined in the second convention $d_{\rm C}(\tilde{P}^{\rm(occ)}_{\bm{k}},\tilde{P}^{\rm(occ)}_{\bm{k}'})$ depends on the choice of the unit cell because of the lack of the information of the intercell positions. Owing to this ambiguity, the distance for some choice of the unit cell can be nonzero for some $\bm{k},\bm{k}'$ even though that for another choice is zero for any $\bm{k},\bm{k}'$.
In the case of the first convention, on the other hand, the eigenstates are not changed by changing the choice of the unit cell, except for overall phase factor, which means that the projection operator is invariant under such a change.  
\section{Definition of nontrivial interband effect \label{sec3}}
In this section, we characterize the nontrivial interband effect as the interband effect that cannot be removed without changing given conditions such as the symmetry.

We first consider the case where there is no interband effect:
\begin{align}
    d_{\rm C}(P^{\rm(occ)}_{\bm{k}},P^{\rm(occ)}_{\bm{k}'})=0\label{dc1}
\end{align}
for any $\bm{k},{\bm{k'}}$. For the second convention of the Fourier transform, 
\begin{align}
    d_{\rm C}(\tilde{P}^{\rm(occ)}_{\bm{k}},\tilde{P}^{\rm(occ)}_{\bm{k}'})=0\label{dc2}
\end{align}
for any $\bm{k},{\bm{k'}}$ and for any choice of the unit cell.
The combination of the conditions (\ref{dc1}) and (\ref{dc2}) is equivalent to the decomposability of the total Bloch Hamiltonian into two independent Bloch Hamiltonians whose basis vectors are characterized by the local atomic orbitals:
\begin{align}
    H_{\bm{k}}=
    \begin{pmatrix}
    H_{\bm{k}}(\mathcal{O})&0\\
    0&H_{\bm{k}}(\overline{\mathcal{O}})
    \end{pmatrix},\label{decomposability}
\end{align}
where $\mathcal{O}$ and $\overline{\mathcal{O}}$ represent the information of the occupied and unoccupied bands. See a proof for Appendix \ref{proof}. 
In real space, the tight-binding Hamiltonian is decomposed into two tight-binding Hamiltonians whose basis vectors are local atomic orbitals:
\begin{align}
    H^{\rm TB}=
    \begin{pmatrix}
    H^{\rm TB}(\mathcal{O})&0\\
    0&H^{\rm TB}(\overline{\mathcal{O}})
    \end{pmatrix}.\label{tbdecomposability}
\end{align}

As discussed above, the absence of the interband effect is equivalent to the decomposability of the tight-binding Hamiltonian into two independent tight-binding Hamiltonians with local basis vectors. In this paper, we define the nontrivial interband effect as the interband effect in which the tight-binding Hamiltonian cannot be expressed as Eq. (\ref{tbdecomposability}) without breaking the $rules$ that one assumes, such as symmetrical and topological constraints. In the following, we formulate the nontrivial interband effect for general cases.  In the next section, we give some examples of rules related to topological properties.

Let $\mathcal{B}$ be some information of the set of considered bands that is changed only when the considered bands tough with other bands, such as the band representation.
We assume that $\mathcal{B}$ belongs to the trivial set $\{\rm trivial \}$ or the nontrivial set $\{\rm nontrivial \}$.
The triviality is defined by using the notion of the tight-binding Hamiltonian.
If one can construct the tight-binding model $H^{\rm TB}(\mathcal{B})$ whose all bands are characterized by $\mathcal{B}$, we say $\mathcal{B}$ is trivial ($\mathcal{B}\in \{\rm trivial \} $). In other words, if one finds a property that should hold in the tight-binding model, then one can define such a property as the triviality.

Now we are in a position to define the nontrivial interband effect.
Suppose that $\mathcal{O}$/$\overline{\mathcal{O}}$ is the information of the occupied/unoccupied band, and the information of the total bands $\mathcal{O}+\overline{\mathcal{O}}$ belongs to the trivial set, which means that the total system can be expressed by the tight-binding Hamiltonian.
Then the following holds by the definition of the triviality:
\begin{align}
    &H^{\rm TB}(\mathcal{O}+\overline{\mathcal{O}})=
    \begin{pmatrix}
    H^{\rm TB}(\mathcal{O})&0\\
    0&H^{\rm TB}(\overline{\mathcal{O}})
    \end{pmatrix}\notag\\
    \Rightarrow &\mathcal{O}\in\{\rm trivial \}{\rm~and~}\overline{\mathcal{O}}\in\{\rm trivial \}.
\end{align}
The left-hand side is equivalent to the condition for the absence of the interband effect between occupied and unoccupied bands.
By considering the contraposition, we obtain 
\begin{align}
    &\mathcal{O}\in\{\rm nontrivial \}{\rm~or~}\overline{\mathcal{O}}\in\{\rm nontrivial \}\notag\\
    \Rightarrow &{\rm There~is~a~nontrivial~interband~effect}.\label{contraposition}
\end{align}

This is the formulation of the nontrivial interband effect.
The type of the information determines that of the nontrivial interband effect, and one can use any information as the triviality/nontriviality as long as it satisfies the assumptions mentioned above.
In the next section, we choose various types of topological concepts as the triviality/nontriviality. Note that the nontrivial interband effect does not require the topological nature in general.
The framework of nontrivial interband effect, which contains topological concepts as subsets, might be useful for finding unexplored concepts.
\section{Nontrivial interband effect and various topological concepts \label{sec4}}
As discussed in the previous section, a nontrivial interband effect is defined by regarding a property of the tight-binding Hamiltonian as the triviality.
In this section, we give examples of nontrivial interband effects that come from the topological band structures. We here omidt the concepts depend on the specific number of bands such as the Hopf insulator \cite{Hopf-ins}. 

\subsection{Nontriviality as stable topology}
A typical example of the information of the set of bands is a topological number defined in a topological insulator, such as the Chern number in the quantum Hall effect \cite{TKNN} and $\mathbb{Z}_2$ invariant in time-reversal-symmetric topological insulators \cite{Kane-review,Zhang-review}.
Since the topological number is changed only when the band inversion occurs, and its summation over all bands of a tight-binding Hamiltonian is zero (trivial), it satisfies the conditions for $\mathcal{B}$ raised in the previous section and characterizes a nontrivial interband effect.
According to (\ref{contraposition}), there is a nontrivial interband effect
if $\mathcal{O}$ or $\overline{\mathcal{O}}$ takes a nonzero topological number, which reproduces the common knowledge that topological band structures have the interband nature. In fact, there should appear the boundary states under the open boundary condition that connect the occupied and unoccupied bands owing to the bulk-boundary correspondence, which can be regarded as a kind of the interband effect.

Note that if one of $\mathcal{O}$ and $\overline{\mathcal{O}}$ is nontrivial, the other one is also nontrivial in the case of the topological insulators. In other words, the addition/subtraction of a trivial element to a nontrivial element becomes always a nontrivial element.
This fact is a consequence of the stable equivalence of the topological K-theory, which classifies the topological insulators and superconductors \cite{Schnyder-08,Kitaev-09,Ryu-10,Schnyder-Ryu-review}.

\subsection{Nontriviality as symmetry-based indicator}
The concept of topological insulators has been generalized to crystalline symmetries, and it is called the topological crystalline insulator (or the higher-order topological insulator in some cases) \cite{LFu-11}.
However, the classification based on the K-theory is not an easy task for general crystalline symmetries and has not been completed despite of several attempts \cite{chiu,morimoto-furusaki,shiozaki-sato-16,okuma-sato-shiozaki-19,cornfeld,shiozaki-19}.
In addition, the explicit construction of the topological invariant such as the TKNN formula \cite{TKNN} is not obtained in the procedures of the K-theoretical classification, and is also a difficult task.
Moreover, the explicit form of a topological invariant contains the momentum integration, which is numerically expensive.

Recently, the concept of the symmetry-based indicator (or related works \cite{kruthoff,chemistry}) has been proposed instead of the K-theoretical framework.
In this framework, the band structure is characterized by the band representations at high-symmetric points in momentum space under the crystalline symmetries.
A ``topologically" trivial band structure is defined as a band structure with the band representation of an atomic insualtor, which is
generated by the localized symmetric Wannier functions. 
If the band structure is not included in the set of the trivial band structures, we say it is ``topologically" nontrivial. The symmetry-based indicator is defined as the set (group) of all band structures divided by that of the trivial band structures defined above.
By construction, there is no guarantee that this ``topology" describes the topological crystalline insulators, and it is known that symmetry-based indicators can classify both the topological (crystalline) insulators and semimetals \cite{po-vishwanath-watanabe,chemistry,diagnosis-1,song-zhang-fang-fang}.
However, the invariants can be computed just from the band representations, and thus this framework is very useful for the realistic material search of topological materials.
One of the earliest examples in this direction is the formula for calculating the parity of the Chern number based on the two-fold rotation eigenvalues \cite{two-fold-formula,any-fold-formula}:
\begin{align}
    (-1)^C=\prod_{a\in\{\rm occ\}}\zeta_a(\Gamma)\zeta_a(X)\zeta_a(Y)\zeta_a(M),\label{two-fold}
\end{align}
where $C$ is the Chern number, $\Gamma,X,Y,M$ are the two-fold rotation-symmetric points in momentum space, and $\zeta_a$ is the two-fold rotation eigenvalue of the band $a$.

In terms of the nontrivial interband effect, the symmetry-based indicator also satisfies the conditions for $\mathcal{B}$ because the total bands of the tight-binding Hamiltonian can always be trivialized to an atomic insulator without changing the indicator by turning off the hopping terms.
Since $\mathcal{O}$ and $\overline{\mathcal{O}}$ can be computed only from the high-symmetric points, one can easily judge the presence of the interband effect.
In this case, the semimetallic band connection or the topological boundary state is the manifestation of the nontrivial interband effect. 
Note that if one of $\mathcal{O}$ and $\overline{\mathcal{O}}$ is nontrivial, the other one is also nontrivial in the case of the symmetry-based indicator, as in the case of the stable topology.
This property comes from the mathematical abstraction that allows the negative numbers of irreducible representations at high-symmetric momenta in order to treat the set of band structures as a group for convenience.

\subsection{Nontriviality as fragile topology}
As we discussed, the symmetry-based indicator enables one to search for topological materials, and it consists of the information of representations at high-symmetric momenta.
However, it does not mean that the band structures with a trivial indicator can be connected to an atomic insulator.
For example, the indicator $(\ref{two-fold})$ of the Chern insulator with even Chern number is trivial, which apparently does not correspond to an atomic insulator.
The fragile topological phase is another case with a trivial indicator \cite{fragile-18}.
Although the band structure does not correspond to topological (crystalline) insulators and semimetals in this phase, it cannot be adiabatically connected to any atomic insulator because the band representation does not correspond to that of any atomic insulator. In contrast to the Chern insulator with even Chern number mentioned above, the fragile topology can still be characterized by the band representation.
The fragile topology corresponds to information of representations that is dropped in the procedure of the mathematical abstraction in the construction of the symmetry-based indicator.

In terms of the nontrivial interband effect,
the band representation itself also satisfies the conditions for $\mathcal{B}$ because the total bands of the tight-binding Hamiltonian can always be trivialized to an atomic insulator without changing the band representation by turning off the hopping terms.
In this case, the fragile topology of $\mathcal{B}$, in addition to the nontrivial symmetry-based indicator, also indicates the presence of the nontrivial interband effect.
In contrast to the stable topology and the symmetry-based indicator, 
 the nontriviality of one of $\mathcal{O}$ and $\overline{\mathcal{O}}$ does not mean the nontriviality of the other one. Thus, the nontrivial band structure can be trivialized by addition of a trivial one, which stems from the fragile topology \cite{fragile-18}.
Interestingly, the interband effect inevitably occurs even though the occupied band is trivial if the unoccupied band is nontrivial, according to Eq. (\ref{contraposition}).

\subsection{Nontriviality as nonzero polarization}
In the case of the symmetry-based indicator and the fragile topology, the atomic insulator is defined by using the localized Wannier functions that are allowed under given crystalline symmetries. 
However, this definition contains the cases where the wave functions are placed not on the atomic positions of the considered material.
This is what happens in the multipole insulator (or some of higher-order topological insulators) \cite{Benalcazar-17}.

Although the multipole insulator is an almost trivial insulator in the sense that the band representation is equal to an atomic insulator, which consists of the direct product of the localized wavefunctions, we can still define the nontrivial interband effect by using the combination of the information of the position of atomic sites and that of the representations.
This is because the band representation tells one the relative positions of the Wannier functions in the unit cell, and one can judge the triviality/nontriviality by the difference between them and the position of the atomic orbitals, called the electric polarization.
Note that if the multipole insulator is exactly an atomic insulator with the flat band, the chordal distance for the second convention of the Fourier transform, $d_{\rm C}(\tilde{P}^{\rm(occ)}_{\bm{k}},\tilde{P}^{\rm(occ)}_{\bm{k}'})$, becomes zero for a choice of the unit cell such that there is no intercell coupling.

\begin{figure*}[]
\begin{center}
　　　\includegraphics[width=16cm,angle=0,clip]{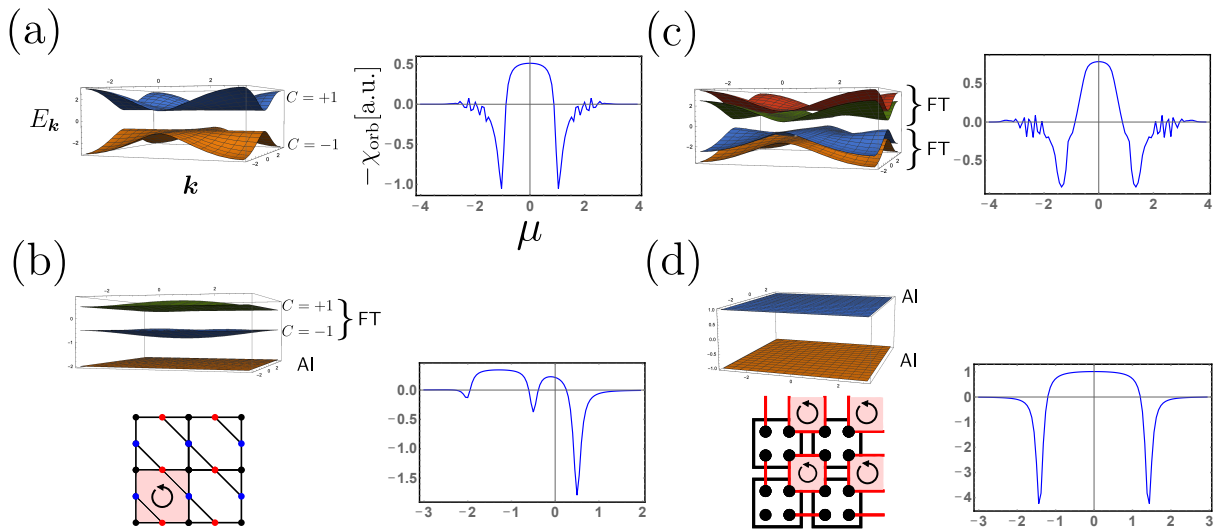}
　　　\caption{Energy dispersion and orbital magnetic susceptibility in topological materials. Each system consists of $32\times32$ unit cells with the periodic boundary condition. The constant self-energy is $\Gamma=0.1$. (a) The model (\ref{Chern}).
　　　The occupied/unoccupied band is described by the Chern number $C=\mp1$.
　　　(b) The model (\ref{fragile}). For an in-gap chemical potential around $\mu=0$, both of the occupied and unoccupied bands are described by odd Chern numbers, which can be detected by the symmetry-based indicator (\ref{two-fold}).
　　　For an in-gap chemical potential around $\mu=-1$, both of the occupied and unoccupied bands are trivial in terms of the symmetry-based indicator (\ref{two-fold}). However, the set of the unoccupied bands hosts the fragile topology (FT), while the occupied band is an atomic insulator (AI) that is not generated on the atomic sites. (c) The model (\ref{frafra}). Occupied/unoccupied bands with fragile topology (FT). (d) The model (\ref{bbh}). Both of the occupied and unoccupied bands are atomic insulators (AIs) that are not generated on the atomic sites.
　}
　　　\label{fig1}
\end{center}
\end{figure*}

\section{Application I: orbital magnetic susceptibility \label{sec5}}
It is known that the interband effect plays an important role in the orbital magnetism.
In this section, we investigate the orbital magnetic susceptibility that originates from the nontrivial interband effect in various topological materials. All of the models are defined on two-dimensional lattices and host the diamagnetism for in-gap chemical potentials.
Although the presence of the nontrivial interband effect is not a sufficient condition for the diamagnetism, it is still useful information for searching the materials with giant diamagnetism.

\subsection{Formula for orbital magnetic susceptibility}
Fukuyama has derived a simple one-line formula for the orbital magnetic susceptibility in terms of Green's functions \cite{fukuyama-71}.
Under the tight-binding approximation, a similar formula has been proposed in Refs.\cite{Koshino-Ando-07,gomez}:  
\begin{align}
    \chi_{\rm orb}=&-\frac{\mu_0e^2}{2\pi\hbar^2}\times{\rm Im}\int_{-\infty}^{\infty}d\omega f(\omega)\notag\\
    &\times\frac{1}{V}\sum_{\bm{k}}\mathrm{Tr}[v_xGv_yGv_xGv_yG\notag\\
    &+\frac{1}{2}(Gv_xGv_y+Gv_yGv_x)G\partial_{k_x}v_y],\label{fukuyama-like}
\end{align}
where $\mu_0$ is the vacuum permeability, $e$ is the elementary charge, $f(\omega)=(e^{(\omega-\mu)/T}+1)^{-1}$ with the chemical potential $\mu$ and the temperature $T$, $V$ is the system volume, $G=(\omega-H_{\bm{k}}+i \Gamma)^{-1}$ is the Green's function with the constant self-energy $\Gamma$, and $v_{x,y}=\partial_{k_{x,y}}H_{\bm{k}}$ is the velocity operator.
Note that this formula is based on the Peierls substitution and ignores several contributions such as from the magnetic-field dependence of the hopping terms \cite{matsuura}. 
In the following, we consider several tight-binding Hamiltonians with topological properties. We
assume that their orbital magnetic susceptibility are well approximated by the formula (\ref{fukuyama-like}).

\subsection{Orbital magnetic susceptibility in topological materials}

We here calculate the orbital magnetic susceptibility of tight-binding Hamiltonians with topological nature by using the formula (\ref{fukuyama-like}). We first consider the Chern insulator \cite{fradkin}:
\begin{align}
    H_{\bm{k}}=\sin k_x \sigma_x+\sin k_y \sigma_y+(m+\cos k_x+\cos k_y)\sigma_z,\label{Chern}
\end{align}
where $\sigma_i$'s are the Pauli matrices, and $m$ denotes the mass.
The Chern number of the occupied band is given by
\begin{align}
C=
    \begin{cases}
    0,~&{\rm for}~m<-2\\
    -1,~&{\rm for}~-2<m<0\\
    +1,~&{\rm for}~0<m<2\\
    0,~&{\rm for}~2<m
    \end{cases}.
\end{align}
The dispersion relation and the orbital magnetic susceptibility with respect to the chemical potential are plotted for $m=-1$ in Fig. \ref{fig1} (a).
For an in-gap chemical potential around $\mu=0$, both of the occupied and unoccupied bands are nontrivial in terms of the stable topology, and the interband effect cannot be removed without closing the gap.
For this parameter region, the orbital magnetic susceptibility takes a negative value, which means that the Chern insulator hosts the diamagnetism. The origin of the diamanetism is the Dirac-like dispersion at the $\Gamma$ point, which is known to play important roles in materials with giant diamagnetism such as bismuth \cite{fkubo-70,fuseya-ogata-15} and graphite/graphene-type compounds \cite{McClure,Koshino-Ando-07,ogata-fukuyama-3}.

Next, we consider the cases with nontrivial indicator and fragile topology by using the following model with three bands $\alpha=1,2,3$ [Fig.\ref{fig1}(b)]:
\begin{align}
    H_{\bm{k}}&=\begin{pmatrix}
    0&\frac{-i}{2}\cos\frac{k_x}{2}&\frac{i}{2}\cos \frac{k_y}{2}\\
    \frac{i}{2}\cos\frac{k_x}{2}&0&\frac{-i}{2}\cos \frac{k_x-k_y}{2}\\
    \frac{-i}{2}\cos \frac{k_y}{2}&\frac{i}{2}\cos \frac{k_x-k_y}{2}&0
    \end{pmatrix}\notag\\
    &-\frac{4}{3+\cos k_x+\cos k_y+\cos (k_x-k_y)}\times\notag\\
    &
    \begin{pmatrix}
    \cos^2\frac{k_x-k_y}{2}&\cos\frac{k_x-k_y}{2}\cos\frac{k_y}{2} &\cos\frac{k_x-k_y}{2}\cos\frac{k_x}{2}\\
    \cos\frac{k_x-k_y}{2}\cos\frac{k_y}{2}&\cos^2\frac{k_y}{2}&\cos\frac{k_x}{2}\cos\frac{k_y}{2}\\
    \cos\frac{k_x-k_y}{2}\cos\frac{k_x}{2}&\cos\frac{k_x}{2}\cos\frac{k_y}{2}&\cos^2\frac{k_x}{2}
    \end{pmatrix}.\label{fragile}
\end{align}
The lowest band is a flat band.
This model has been investigated in Ref. \cite{dominic-19} in the second convention of the Fourier transform. 
In terms of the stable topology, the lowest band ($\alpha=1$) is trivial, while the upper two bands ($\alpha=2,3$) are nontrivial because of the odd Chern numbers.
The nontriviality can also be detected by the symmetry-based indicator (\ref{two-fold}).
The system has the spinless two-fold rotation symmetry (or inversion symmetry in two dimensions), and two-fold rotation eigenvalues (parity) at two-fold-rotation-symmetric points are given by (see Ref. \cite{dominic-19} for details)
\begin{align}
    \bm{\zeta}_1&=(+1,-1,-1,+1),\notag\\
    \bm{\zeta}_2&=(+1,+1,+1,-1),\notag\\
    \bm{\zeta}_3&=(+1,+1,+1,-1),
\end{align}
where $\bm{\zeta}_\alpha:=(\zeta_\alpha(\Gamma),\zeta_\alpha(X),\zeta_\alpha(Y),\zeta_\alpha(M))$. 
According to Eq. (\ref{two-fold}), the lowest band $\alpha=1$ is trivial, while the upper two bands $\alpha=2,3$ are nontrivial.
Thus, for an in-gap chemical potential around $\mu=0$, both of the occupied and unoccupied bands are nontrivial in terms of the symmetry-based indicators, and the interband effect cannot be removed without closing the gap. For this parameter region, the system hosts the diamagnetism.

The fragile topology is another interesting perspective in this model \cite{dominic-19}.
For the upper two bands $\alpha=2,3$,
both of the total Chern number and the symmetry-based indicator (\ref{two-fold}) are trivial.
However, the representation $\bm{\zeta}_{2}\oplus\bm{\zeta}_{3}$ does not correspond to any atomic insulator for two-fold rotation symmetry, which means that it hosts the fragile topology. Although the lowest band $\alpha=1$ is an atomic insulator, equation (\ref{contraposition}) indicates that there is a nontrivial interband effect between the lowest band and the set of the upper bands.
For an in-gap chemical potential around $\mu=-1.5$, the system hosts the diamagnetism. Note that the occupied atomic insulator is generated not on the atomic sites but on a two-fold-rotation symmetric intermediate region in the unit cell \cite{dominic-19}, and the wavefunction consists of ``bonding orbitals" of the atomic orbitals.
Roughly speaking, this diamagnetism can be regarded as consequence of the ring current on these ``bonding orbitals".

Both of the occupied and unoccupied bands of following model can host the fragile topology:
\begin{align}
    H_{\bm{k}}=&\sin k_x \sigma_x\otimes \tau_z+\sin k_y \sigma_y\otimes\tau_0\notag\\
    &+(m+\cos k_x+\cos k_y)\sigma_z\otimes \tau_0+\delta \sigma_0\otimes\tau_x,\label{frafra}
\end{align}
where $\sigma_\mu$'s and $\tau_\mu$'s are the Pauli matrices. This model consists of two Chern insulators with opposite Chern number with a mixing term $\delta$ between them.
This Hamiltonian has the two-fold rotation symmetry:
\begin{align}
    C_2H_{\bm{k}}C_2^{-1}=H_{-\bm{k}},
\end{align}
where $C_2=\sigma_z\otimes \tau_0$.
The dispersion relation and the orbital magnetic susceptibility with respect to the chemical potential are plotted for $m=-1$ and $\delta=0.5$ in Fig. \ref{fig1} (c).
The two-fold rotation eigenvalues (parity) of the occupied $\alpha=1,2$ and the unoccupied bands $\alpha=3,4$ at two-fold-rotation-symmetric points are given by
\begin{align}
    \bm{\zeta}_1&=(-1,+1,+1,+1),\notag\\
    \bm{\zeta}_2&=(-1,+1,+1,+1),\notag\\
    \bm{\zeta}_3&=(+1,-1,-1,-1),\notag\\
    \bm{\zeta}_4&=(+1,-1,-1,-1).
\end{align}
According to Eq. (\ref{two-fold}), both of the occupied and unoccupied bands are trivial. However, both of them cannot be connected to atomic insulators and host fragile topology.
For an in-gap region around $\mu=0$, the system hosts the diamagnetism.

Finally, we consider the Benalcazar-Bernevig-Hughes (BBH) model \cite{Benalcazar-17}, which is a prototypical example of the multipole insulator (or  higher-order topological insulator):
\begin{align}
    H_{\bm{k}}&=(\gamma+\lambda \cos k_x) \tau_1\otimes\sigma_0-(\gamma+\lambda \cos k_y)\tau_2\otimes\sigma_2 \notag\\
    &-\lambda\sin k_x \tau_2\otimes\sigma_3-\lambda\sin k_y \tau_2\otimes\sigma_1+\delta \tau_3\otimes\sigma_0.\label{bbh}
\end{align}
We assume that the four internal degrees of freedom are located at the same place in the unit cell for simplicity.
For $\delta=0$, the BBH model has the four-fold rotation symmetry:
\begin{align}
    &C_4H_{\bm{k}}C_4^\dagger=H_{C_4\bm{k}},\notag\\
    &C_4=
    \begin{pmatrix}
    0&\sigma_0\\
    -i\sigma_y&0
    \end{pmatrix}.
\end{align}
The dispersion relation and the orbital magnetic susceptibility with respect to the chemical potential are plotted for $\delta=0$ and $\gamma=0$ in Fig. \ref{fig1} (d).
Both of the occupied and unoccupied bands are atomic insulators generated not on the atomic sites but on the four-fold symmetric position. In other words, this system is a multipole insulator. This can be checked by the indicator-type formula in Refs. \cite{corner-19,corner-19-2}. 
Owing to the multipole moment, the interband effect cannot be removed without closing the gap or the breaking the symmetry. 
For an in-gap region around $\mu=0$, the system has the diamagnetism.
Since the system at the present parameters can be decomposed into an infinite number of tetramers, this diamagnetism has the same origin as the diamagnetism of molecules, such as benzene, induced by the ring current \cite{fujii-51,matsuura}.

\section{Application II: interband-induced degeneracy pressure \label{sec6}}
In this section, we consider the degeneracy pressure from the free electrons in terms of the interband effect. We first review the conventional degeneracy pressure in the Fermi gas \cite{Fetter}, which is the simplest model of a metal. We then consider the interband-induced degeneracy pressure in insulators and semimetals without the finite Fermi surfaces.
\subsection{Degeneracy pressure in Fermi gas}
In the Fermi gas \cite{Fetter}, the Fermi energy and the total energy are given by
\begin{align}
    &E_F=\frac{1}{2m}\left(\frac{3\pi^2 N}{V}\right)^{\frac{2}{3}},\notag\\
    &E_{\rm tot}=V\int_0^{E_F}dE~E~g(E)=\frac{3}{5}NE_F,
\end{align}
where $m$ is the electron mass, $N$ is the number of electrons, $g(E)=m\sqrt{2mE}/\pi^2$ is the density of states, and $V$ is the system volume.
Then the pressure defined for the Fermi gas is given by
\begin{align}
    p:=-\frac{d E_{\rm tot}}{d V}=\frac{2}{3}\frac{E_{\rm tot}}{V}>0.
\end{align}
This positive pressure is called the electron degeneracy pressure, which exists even at the zero temperature. This degeneracy pressure defines the inverse compressibility, or the bulk modulus:
\begin{align}
    B:=-V\frac{d p}{d V}=\frac{10}{9}\frac{E_{\rm tot}}{V}>0.
\end{align}
In terms of the band theory, these quantities come from the intraband effect.
In the following, we consider the interband analogue of the degeneracy pressure.
\subsection{Interband-induced degeneracy pressure}
We here consider the interband-induced degeneracy pressure of the tight-binding Hamiltonians defined for free fermions.
We assume that Tr $[P^{\rm(occ)}_{\bm{k}}]$ is independent of $\bm{k}$, which means that the system is an insulator or a semimetal without the finite Fermi surface.
The degeneracy pressure, or internal pressure, is defined by
\begin{align}
    p:=&-\frac{d}{dV}~\mathrm{Tr}~[P^{\rm(occ)}_{\rm TB}H^{\rm TB}]\notag\\
    =&-\frac{d}{dV}\left(\sum_{\alpha\in\{\rm occ\}}\langle \alpha|H^{\rm TB}|\alpha\rangle \right)\notag\\
    =&-\sum_{\alpha\in\{\rm occ\}}\langle \alpha|\frac{dH^{\rm TB}}{dV}|\alpha\rangle\notag\\
    =&-{\rm Tr}~\left[P^{\rm(occ)}_{\rm TB}\frac{dH^{\rm TB}}{dV}\right],\label{defofpressure}
\end{align}
where $|\alpha\rangle$'s are the eigenstates of $H^{\rm TB}$, and $P^{\rm(occ)}_{\rm TB}$ is the real-space representation of the projection operator onto the occupied bands. On the third line, we have used $d(\langle \alpha|\alpha\rangle)/dV=0$.
For the degeneracy pressure, one can define the corresponding inverse compressibility:
\begin{align}
    B:=-V\frac{d p}{d V}=V{\rm Tr}~\left[P^{\rm(occ)}_{\rm TB}\frac{d^2H^{\rm TB}}{dV^2}\right].
\end{align}

The explicit calculation of  Eq.(\ref{defofpressure}) cannot be computed without the information of the volume-dependence of the matrix elements of $H^{\rm TB}$.
We further assume that the onsite elements are independent of the volume change.
This assumption might be reasonable at least for a small volume change that does not affect the shape of the local atomic orbitals.
Under this assumption, the following holds for general tight-binding Hamiltonians:
\begin{align}
    \mathrm{Tr}~\left[\frac{dH^{\rm TB}}{dV}\right]=\frac{d}{dV}(\sum_iH_{ii})=0.\label{trivpressure}
\end{align}
If there is no interband effect between the occupied and unoccupied bands, one can construct the tight-binding Hamiltonian of the occupied bands.
Thus, the absence of the interband effect indicates that the degeneracy pressure (and inverse compressibility) is 0, according to Eq. (\ref{trivpressure}).
By contraposition, nonzero degeneracy pressure indicates the presence of the interband effect under the assumptions above. Note that the metallic degeneracy pressure treated in the previous subsection is absent because of the assumption that Tr $[P^{\rm(occ)}_{\bm{k}}]$ is independent of $\bm{k}$.
In general, both of the intraband and interband contributions can exist in metallic materials.

\subsection{Negative degeneracy pressure}
To proceed further, we consider the cases in which all of the amplitude of the inter-atomic hopping elements decrease with the same rate for the increase of the volume:
\begin{align}
    \frac{dH^{\rm TB}}{dV}=-r(V)(H^{\rm TB}-H^{\rm O}), \label{bane}
\end{align}
where $r(V)$ is a positive value, and $H^{\rm O}$ is the onsite part of $H^{\rm TB}$.
Under this assumption, the degeneracy pressure is shown to be less than 0, at least for the case with $D_{\bm{G}_0}=1$ with $\bm{G}_0$'s being the unit reciprocal lattice vectors:
\begin{align}
    p&=-{\rm Tr}~\left[P^{\rm(occ)}_{\rm TB}\frac{dH^{\rm TB}}{dV}\right]
    =-\sum_{\bm{k}}{\rm Tr}~\left[P^{\rm(occ)}_{\bm{k}}\frac{dH_{\bm{k}}}{dV}\right]\notag\\
    &=r \sum_{\bm{k}}{\rm Tr}~\left[P^{\rm(occ)}_{\bm{k}}H_{\bm{k}}-P^{\rm(occ)}_{\bm{k}}H^{\rm O}\right]\notag\\
    &=r\sum_{\bm{k}}\left({\rm Tr}~\left[P^{\rm(occ)}_{\bm{k}}H_{\bm{k}}\right]-\frac{1}{N_u}\sum_{\bm{k}'}{\rm Tr}~\left[P^{\rm(occ)}_{\bm{k}}H_{\bm{k}'}\right]        \right)\notag\\
    &\leq r\left(\sum_{\bm{k}}{\rm Tr}~\left[P^{\rm(occ)}_{\bm{k}}H_{\bm{k}}\right]-\sum_{\bm{k}'}\left[P^{\rm(occ)}_{\bm{k}'}H_{\bm{k}'}\right]\right)=0,\label{result}
\end{align}
where $N_u$ is the number of unit cells.
On the second line, we have used the notation $H^{\rm O}$ for the Fourier transform because of the absence of the $\bm{k}$-dependence.
On the third line, we have used 
\begin{align}
    H^{\rm O}=\frac{1}{N_u}\sum_{\bm{k}'}H_{\bm{k}'},\label{hon}
\end{align}
which holds for periodic $H_{\bm{k}}$ ($D_{\bm{G}_0}=1$).
On the fourth line, we have used 
\begin{align}
    {\rm Tr}~\left[P^{\rm(occ)}_{\bm{k}}H_{\bm{k}'}\right]\geq{\rm Tr}~\left[P^{\rm(occ)}_{\bm{k}'}H_{\bm{k}'}\right].\label{ineq}
\end{align}
This holds because the right-hand side gives the ground state energy of the subsystem with particle number $\mathrm{Tr}~[P^{\rm(occ)}_{\bm{k}'}]$ whose Hamiltonian is $H_{\bm{k}'}$. The equality holds when $P^{\rm(occ)}_{\bm{k}}$ is independent of $\bm{k}$.
Since the first and second conventions are equivalent for $D_{\bm{G}_0}=1$, Eqs. (\ref{dc1},\ref{dc2}) hold, and the condition for the equality is equivalent to the absence of the interband effect.

For the cases with $D_{\bm{G}_0}\neq1$, Eq. (\ref{ineq}) does not hold because the cancellation of the $e^{i\bm{k}\cdot\bm{r}_i}$ does not occur.
If there is a set of $\bm{G}$'s such that $D_{\bm{G}}=1$, the same proof can be used by extending the Brillouin zone in Eq. (\ref{hon}) and using the fact that ${\rm Tr}~\left[P^{\rm(occ)}_{\bm{k}'}H_{\bm{k}'}\right]$ is always periodic in $\bm{G}_0$'s.
In the cases without such $\bm{G}$'s, the above proof should be modified. However, one might be able to find a set of $\bm{G}$'s such that $D_{\bm{G}}$ are approximately an identity matrix, and the error could be taken as an arbitrary small value, which controls the right-hand side of Eq. (\ref{result}).

\begin{figure}[]
\begin{center}
　　　\includegraphics[width=8cm,angle=0,clip]{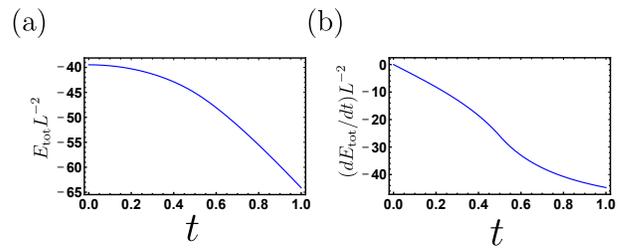}
　　　\caption{(a) Total-energy per unit cell and (b) its $t$-derivative of the model ($\ref{chern2}$).
　}
　　　\label{fig2}
\end{center}
\end{figure}

When all of the above assumptions are satisfied, the nontrivial interband effect indicates the negative degeneracy pressure.
For example, let us again consider the Chern insulator:
\begin{align}
    H_{\bm{k}}=t\sin k_x \sigma_x+t\sin k_y \sigma_y+(-1+t\cos k_x+t\cos k_y)\sigma_z,\label{chern2}
\end{align}
where $t$ is introduced to connect the trivial ($t=0$) and the nontrivial ($t=1$) phases. The gap-closing point is at $t=0.5$.
The total energy per unit cell and its derivative are plotted in Fig. \ref{fig2}.
At $t=1$, $dE_{\rm tot}/dt$ is negative, while it is zero at $t=0$.
Since the pressure is given by
\begin{align}
    p=-\frac{dt}{dV}\frac{dE_{\rm tot}}{dt},
\end{align}
the sign of $dt/dV$ determines the sign of the pressure. When we assume that the hopping term $t$ is a monotonically decreasing function of $V$, the degeneracy pressure takes a negative value at $t=1$. Because the physical origin of the hopping is the transfer integral between the neighbor atomic orbitals with overlap, this is a reasonable assumption.

In summary, we have shown that the degeneracy pressure becomes negative in the presence of the interband effect under the above assumptions.
Although the assumption ($\ref{bane}$) seems to be too strong to be realized in general materials, a similar statement could hold for a different form of the right-hand side.
For example, if we can find
\begin{align}
    H^{\rm O}\propto\sum_{\bm{k}'}U_{\bm{k}'}H_{\bm{k}'}U_{\bm{k}'}^{\dagger},
\end{align}
with $U$ being a unitary matrix, we could still use a similar proof. 
A typical example of such a case is the unitary transformation from the first convention to the second convention of the Fourier transform.
In this sense, we expect that the degeneracy pressure induced by the interband effect tends to be negative in realistic materials.
The nontrivial interband effects in quantum materials such as topological insulators play important roles not only in the electromagnetic responses but also in the mechanical responses such as the degeneracy pressure and the inverse compressibility.
The negative degeneracy pressure behaves as if an external pressure to the system, which might be useful for quantum material science that requires the high pressure, such as the physics of high-temperature superconductors \cite{Drozdov}.
The competition between the intraband and interband effects in metals is also an important remaining topics.

\begin{figure}[]
\begin{center}
　　　\includegraphics[width=8cm,angle=0,clip]{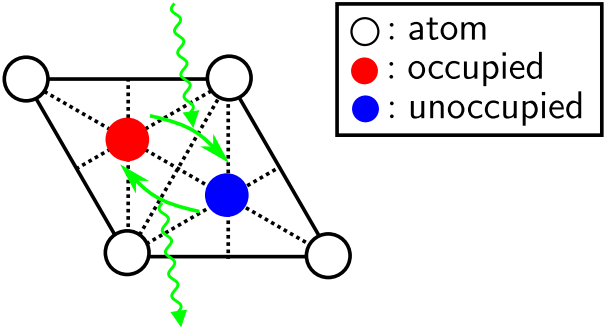}
　　　\caption{Schematic picture of optical response in unit cell of a multipole insulator under $C_3$ and time-reversal symmetries. The quantized polarization of the occupied bands indicates the presense of the nontrivial interband effect and the nonzero difference between the polarization of the occupied and that of the unoccupied bands. This polarization difference can contribute to the shift current (\ref{shift}).
　}
　　　\label{fig3}
\end{center}
\end{figure}

\section{Application III: nonlinear optics\label{sec7}}
Nonlinear optics is another important example related to the interband effect.
Reference \cite{Morimoto-nagaosa} has shown the topological nature of the nonlinear optics. 
For example, the optically induced shift current in the second harmonic generation for a two-band approximation is given by \cite{Morimoto-nagaosa}
\begin{align}
    J(2\Omega)\simeq &\frac{\pi E^2}{2\Omega^2}\int dk |v_{12}|^2R_k\notag\\
    &\left[-\delta(\epsilon_1-\epsilon_2+\Omega)+\frac{1}{2}\delta(\epsilon_1-\epsilon_2+2\Omega)\right],\label{shift}
\end{align}
where $\Omega$ is the frequency of the light, $E$ is the amplitude of the electric field of the light, $\epsilon_1/\epsilon_2$ is the dispersion of the occupied ($\alpha=1$)/unoccupied ($\alpha=2$) states, $v_{12}$ is the matrix element of the velocity operator between the occupied and unoccupied bands, and 
\begin{align}
    R_{k}=\frac{\partial \mathrm{Im}(\mathrm{log} v_{12})}{\partial k}+\frac{1}{i}\langle u_1|\partial_k|u_1\rangle-\frac{1}{i}\langle u_2|\partial_k|u_2\rangle
\end{align}
is the shift vector between the occupied and the unoccupied eigenstates $|u_1\rangle,|u_2\rangle$. The shift vector measures the difference of the intracell polarizations between two bands. According to this formula, the interband effect between optically-coupled bands ($v_{12}\neq0$) and the polarization difference ($R_k$) are essential factors for the shift current. 
For material search in this area,
the concept of the nontrivial interband effect might be useful because the presence of the interband effect and the difference of the polarizations can be judged by the band information such as the representations.
%For example, let us consider a two-dimensional tight-binding model with $C_{3v}$ symmetry, without time-reversal symmetry. Suppose that the basis vectors of the Bloch Hamiltonian are characterized by six atomic orbitals that sit on the nonsymmetric points (Wyckoff position with multiplicity 6). If the band representations of the occupied and unoccupied bands are those of the atomic insulators generated on the different $C_{3v}$-symmetric points (Wykoff positions with multiplicity 1), then the interband effect inevitably occurs, and the difference of the polarizations is nonzero.

As an example, let us consider a multipole (higher-order topological) insulator with $C_3$ and time-reversal symmetries. According to Refs. \cite{corner-19,corner-19-2}, the quantized polarization can be determined by an indicator ($\chi_3$). If the total bands are well described by a corresponding tight-binding model with no polarization, the nontrivial polarization $\bm{P}=(e,e)/3$ or $(-e,-e)/3$ of the occupied bands indicates the nontrivial polarization $\bm{P}=(-e,-e)/3$ or $(e,e)/3$ of the unoccupied bands, which means that there are the nontrivial interband effect and the polarization difference $\Delta \bm{P}=\pm(e,e)/3$ (Fig.\ref{fig3}).
This fractional difference can contribute to the shift current (\ref{shift}).
Note that in two dimensions, the polarization difference $\Delta \bm{P}$ is not fractional for $C_2$, $C_4$, and $C_6$ symmetries, which contain the two-fold rotation symmetry (or inversion symmetry in two dimensions). 

A fragile topological insulator is another candidate for the quantized polarization difference.
As mentioned above, the fragile topology can be trivialized by an atomic insulator. As mentioned in Ref. \cite{dominic-19}, such an atomic insulator is generated not on atomic sites, which means that it has a nontrivial polarization. In other words, the fragile topology is also characterized by a nontrivial electronic polarization with the opposite value. Thus, the above discussion about the polarization difference for the multipole insulators can also be applied to the fragile topological insulators.

The relationship between the fractional polarization difference with the nontrivial interband effect and the indicator-type formula for general symmetries and dimensions is an important remaining issue.
In the field of nonlinear optics, some of the symmetries of solids such as the inversion symmetry are known to forbid or give constraints on the nonlinear optical response. As well as such traditional information, the band representation with a nontrivial interband effect  can be regarded as useful information for material search.

\section{Generalization to interacting system with entanglement \label{sec8}}
In interacting systems, the band picture cannot be naively introduced.
One possible way to generalize the interband effect is to introduce the flux insertion to the periodic boundary condition \cite{oshikawa}:
\begin{align}
    H_{\bm{\theta}}|GS,\bm{\theta}\rangle&=E_{\bm{\theta}}|GS,\bm{\theta}\rangle,\notag\\
    H_{(\cdots,\theta_i,\cdots)}&=U_iH_{(\cdots,\theta_i+2\pi,\cdots)}U^{\dagger}_i,
\end{align}
where $H$ is a many-body Hamiltonian, $\theta_i$ is the flux through the ring in the $i$ direction, $GS$ denotes the ground state(s), and $U_i$'s are unitary operators called twist operators. Physically, the phase factor $e^{i\theta/L}$ is assigned at each bond by the flux insertion. The projection operator is given by
\begin{align}
    P_{\bm{\theta}}=\sum_{GS}|GS,\theta\rangle\langle GS,\theta|.
\end{align}
By defining the chordal distance for $P_{\bm{\theta}}$, most of the discussions in the previous sections can formally be generalized to interacting cases.
Among them, we here consider the degeneracy pressure:
\begin{align}
    p:=-{\rm Tr}~\left[P\frac{dH}{dV}\right].
\end{align}
By considering the counterpart of the assumption ($\ref{bane}$), we again obtain the negative degeneracy pressure. Here we have assumed that $U_i^{L_i}=1$ with the system size in the $i$ direction $L_i$ and used
\begin{align}
    H^{\rm O}=\prod_i\int^{2\pi L_i}_{0} \frac{d^d\theta}{(2\pi)^d}H_{\bm{\theta}},
\end{align}
where $d$ is the space dimension. 
Note that one can choose arbitrary phase factors whose corresponding Hamiltonian is unitary-equivalent to $H_{\bm{\theta}}$.
This property enables one to choose variants of $H^{\rm O}$ for the proof, which depends on the gauge choice in general. Thus, the negativity of the degeneracy pressure holds at least for $H^{\rm O}$ and its variants.
Off course, all of the above discussions can also be applied to non-interacting systems.

By construction under the assumption that only the inter-site interactions depend on the volume, the negative pressure detects the entanglement of the bonds.
For example, let us consider the Affleck-Kennedy-Lieb-Tasaki model \cite{aklt}:
\begin{align}
    H_{\rm AKLT}=J\sum_i\left[\bm{S}_i\cdot\bm{S}_{i+1}+\frac{1}{3}(\bm{S}_i\cdot\bm{S}_{i+1})^2 \right],
\end{align}
where $J>0$, $i$ denotes a site, and $\bm{S}_i$ is the spin-1 operator at site $i$. This model has a unique gapped ground state written in a matrix product state, which is a typical example of the short-range entangled state.
The ground-state energy is simply given by $-2JL/3$ with $L$ being the system size. When we assume that $J$ is a monotonically decreasing function of the volume, the degenerate pressure takes a negative value. In Kitaev's toric code model \cite{kitaev-03}, the ground state(s) hosts the long-range entanglement, and the ground-state energy also depends on the bond coupling constants, which means the presence of the degenerate pressure.
In this sense, the interband effect in the noninteracting systems and the entanglement in general interacting systems including the symmetry-protected/intrinsic topological orders are closely related to each other.
Thus, some of interband-induced phenomena could be defined for the ground states with entanglement of interacting systems. 
Note that we roughly use the terminology ``degeneracy pressure" for spin systems in the sense that the filling is fixed.

\section{Summary \label{sec9}}
In this paper, we have presented a theory of the nontrivial interband effect, which cannot be removed without breaking given conditions.
We have defined the general nontrivial interband effect by regarding a property of the set of the total bands of a tight-binding Hamiltonian as the triviality.
As examples of the source of the nontriviality, we have considered several topological concepts: stable topology,  symmetry-based indicator, fragile topology, and quantized multipole moment.
Since the interband effect plays important roles in solid-state physics, the concept of the nontrivial interband effect is useful for material search.
As an application, we have calculated interband-induced orbital magnetic susceptibility for topological materials.
In addition, we have proposed the notion of the interband-induced degeneracy pressure, which tends to take a negative value.
We have also discussed the application to nonlinear optics and the generalization to interacting systems with entanglement.

\acknowledgements
N.O. thanks all the people in Masao Ogata's group for fruitful discussions about the diamagnetism during my Ph.D course. 
This work was supported by JST CREST Grant No.~JPMJCR19T2, Japan. N.O. was supported by KAKENHI Grant No.~JP20K14373 from the JSPS.

\appendix
\section{Proof of Eq. (\ref{decomposability})$\Leftrightarrow$ Eqs. (\ref{dc1}) and (\ref{dc2}) \label{proof}}
We here show that Eq. (\ref{decomposability})$\Leftrightarrow$ Eqs. (\ref{dc1}) and (\ref{dc2}).
\\
\\
\underline{Eq. (\ref{decomposability})$\Rightarrow$ Eqs. (\ref{dc1}) and (\ref{dc2})}\\
``Eq. (\ref{decomposability})$\Rightarrow$ Eq. (\ref{dc1})" is trivial because the basis vectors are fixed with respect to the momentum, and $P_{\bm{k}}=$ diag($1,1,\cdots,1,0,\cdots,0$) in the common basis vectors.
Since $D_{\bm{k}}$ in Eq. (\ref{firstsecond}) is diagonal in the local basis vectors, the Bloch Hamiltonian in the second convention can also be decomposed into two independent Bloch Hamiltonian with local atomic orbitals:
\begin{align}
    \tilde{H}_{\bm{k}}=
    \begin{pmatrix}
    \tilde{H}_{\bm{k}}(\mathcal{O})&0\\
    0&\tilde{H}_{\bm{k}}(\overline{\mathcal{O}})
    \end{pmatrix}.
\end{align}
Thus, ``Eq. (\ref{decomposability})$\Rightarrow$ Eq. (\ref{dc2})" should hold for the same reason as discussed above. 
\\
\\
\underline{Eqs. (\ref{dc1}) and (\ref{dc2})$\Rightarrow$ Eq. (\ref{decomposability})}\\
From Eq. (\ref{dc1}), we obtain
\begin{align}
    P_{\bm{k}}=P_{\bm{k}'}=P_0,\label{const}
\end{align}
where $P_0$ is a constant projection matrix.
In other words, one can choose the momentum-independent basis vectors in which the total Bloch Hamiltonian $H_{\bm{k}}$ is decomposed into two Bloch Hamiltonians. Next, we show the locality of the basis vectors. 
From Eqs. (\ref{dc2}) and (\ref{const}), we obtain
\begin{align}
    &\tilde{P}_{\bm{k}}=\tilde{P}_{\bm{k}'}\notag\\
    \Leftrightarrow~ &D^\dagger_{\bm{k}}P_0D_{\bm{k}}= D^\dagger_{\bm{k}'}P_0D_{\bm{k}'}\notag\\
    \Leftrightarrow~ &[P_0,D_{\bm{k}-\bm{k}'}]=0,
\end{align}
where we have used the expression $D_{\bm{k}}=$ diag$(\cdots,e^{-i\bm{k}\cdot\bm{r}_i},\cdots)$.
Thus, $P_k=P_0$ and $D_{\bm{k}}$ can be simultaneously diagonalized.
Since the eigenstates of $D_{\bm{k}}$ are nothing but the local atomic orbitals that span the Hilbert space of the total system, the basis vectors of the occupied/unoccupied Bloch Hamiltonian can be taken as the local ones.

%\bibliography{topo}

\end{document}